**Photo-induced Doping in Graphene/Boron Nitride Heterostructures**

L. Ju[1]*, J. Velasco Jr.[1]*, E. Huang[1], S. Kahn[1], C. Nosiglia[1], Hsin-Zon Tsai[1], W. Yang[2], T. Taniguchi[3], K. Watanabe[3], Y. Zhang[4], G. Zhang[2], M. Crommie[1,5], A. Zettl[1,5], F. Wang[1,5]†

[1]Department of Physics, University of California, Berkeley, California 94720, USA

[2]Beijing National Laboratory for Condensed Matter Physics and Institute of Physics, Chinese Academy of Sciences, Beijing 100190, China

[3]National Institute for Materials Science, 1-1 Namiki, Tsukuba, 305-0044, Japan

[4]State Key Laboratory of Surface Physics and Department of Physics, Fudan University, Shanghai 200433, China

[5]Materials Sciences Division, Lawrence Berkeley National Laboratory, Berkeley, California 94720, USA

*These authors contribute equally to this manuscript.
† Email:fengwang76@berkeley.edu

**The design of stacks of layered materials in which adjacent layers interact by van der Waals forces[1] has enabled the combination of various two-dimensional crystals with different electrical, optical and mechanical properties, and the emergence of novel physical phenomena and device functionality[2-8]. Here we report photo-induced doping in van der Waals heterostructures (VDHs) consisting of graphene and boron nitride layers. It enables flexible and repeatable writing and erasing of charge doping in graphene with visible light. We demonstrate that this photo-induced doping maintains the high carrier mobility of the graphene-boron nitride (G/BN) heterostructure, which resembles the modulation doping technique used in semiconductor heterojunctions, and can be used to generate spatially-varying doping profiles such as *p-n* junctions. We show that this photo-induced doping arises from microscopically coupled optical and electrical responses of G/BN**

**heterostructures, which includes optical excitation of defect transitions in boron nitride, electrical transport in graphene, and charge transfer between boron nitride and graphene.**

Recent scanning probe and electrical transport studies have revealed moiré patterns[2-4], new Dirac points[5], and the Hofstadter butterfly[6-8] in G/BN heterostructures, hence convincingly demonstrating that the interaction between the constituents of VDHs plays a key role in their properties. Light-matter interactions in the VDHs can also exhibit new phenomena from the coupling between the layered constituents. Incidentally, a recent work exploited the strong optical absorption of $WS_2$ and the tunability of graphene electrodes to engineer graphene-$WS_2$-graphene heterostructures for flexible photovoltaic devices with high quantum efficiency[9]. Here we demonstrate an emerging optoelectronic response in G/BN heterostructures, where patterned doping of graphene can be controllably written and erased through optical excitation of BN. This photo-induced doping is analogous to the modulation doping in semiconductor heterojunctions in that it preserves the remarkably high mobility of G/BN[10,11] by having the dopants separated from the conducting channel. This photo-induced "modulation" doping in G/BN heterostructures arises from defect states in the bulk of crystalline BN flakes. It is qualitatively different from photo-induced effects previously observed in graphene on $SiO_2$ (G/$SiO_2$) that originates from interfacial charge traps in the amorphous oxide[12,13]. Moreover, compared to G/$SiO_2$ devices, the photo-doping response in G/BN is more than 1000 times stronger; it leads to an electron mobility more than an order of magnitude higher; and it has superior stability and reversibility. These unique features of photo-doping in G/BN could enable novel high-quality graphene electronic devices using a photoresist-free photolithography, where the BN substrate itself acts as the photosensitive media.

Graphene charge transport in the G/BN heterostructure can be modified upon illumination even with an incandescent lamp, as we show in Fig. 1. We monitor the graphene resistance $R$ while sweeping the bottom gate voltage $V_g$ with different optical excitation conditions. The graphene resistance shows a sharp peak at the charge neutral point (CNP) with $V_g \sim 0$V and has little hysteresis in the absence of light exposure (Fig. 1a), similar to that observed in many graphene field-effect-transistor devices[10,11,14,15]. The transport behavior becomes completely different, however, when the device is exposed to moderate optical illumination.

In Fig. 1b, we plot the gate-dependent graphene resistance when the device is under the illumination of an incandescent lamp. $V_g$ is swept from 70V to -70V and then back to 70V. We observe that $R$ increases initially until it reaches the CNP at $V_g$=0V. Afterwards, $R$ remains constant when $V_g$ sweeps to negative voltages, as if the gating is not working. This is in striking contrast to the behavior without light illumination (Fig. 1a). When the $V_g$ sweep is reversed, $R$ drops quickly as if the reversal point (rather than $V_g$=0) is the CNP.

To further investigate this intriguing photo-induced response, we measured $R(V_g)$ with the light switched off and on alternately as $V_g$ sweeps from 20V to -50V at a constant rate of 0.05V/s (Fig. 1c). We observed conventional $R(V_g)$ behaviour when the light is off (with $V_g$= 20V to -5V), and $R$ increases quickly and becomes pinned at the CNP value whenever light is switched on. Once the light is switched off, $R$ decreases from the CNP value with more negative $V_g$.

Although the photo-induced responses at different illumination conditions seem quite complicated, they can be understood with a very simple phenomenological model illustrated in

Fig. 1d and 1e. In this model, visible light induces a positive charge distribution in BN that completely screens the negative $V_g$, and the positive charges in BN are fixed when the light is switched off. This model explains qualitatively all of our experimental observations: The backgate is screened and graphene is pinned at CNP with light on and a negative $V_g$, and this $V_g$ sets the CNP when light is off or $V_g$ becomes less negative. When the $V_g$ is removed, it results in a stable n-type doping in graphene. Because the positive dopants (in BN) are away from the n-type conducting channel (in graphene), this is analogous to modulation doping first developed for high quality semiconductor heterojunctions[16,17], where dopants are separated from the conducting channel to prevent charge scattering. Although the photo-induced "modulation" doping in G/BN is not permanent, we found it can last for many days at room temperature when the device is kept in a dark environment. Additionally, it can be easily erased in minutes with light illumination at an intensity ~10μW/μm$^2$. Further measurements show that a p-type doping can also be induced in G/BN by optical excitation at a positive $V_g$, but the dynamics is orders of magnitude slower. We have measured 27 G/BN samples and all exhibit the photo-doping effect. However, the exact dynamics of photo-doping can vary significantly upon both the BN flake thickness and the batch of the parent BN crystals.

The photo-induced modulation doping offers two distinct advantages for novel graphene electronic and opto-electronic devices. First, optical illumination gives incredible flexibility to control the doping: different doping concentrations and patterns can be written using light, and they can be generated and erased at will. Secondly, the doping mechanism preserves the remarkably high mobility typical of G/BN.

We show excellent transport properties of graphene with photo-induced modulation doping in Fig. 2a and 2b. We control the doping level by exposing the device to light at fixed $V_g$ until the resistance stabilises and then take an $R(V_g)$ scan with the light off. We used $V_g$ set points at 0(before doping),-10,-20,-30,-40 and -50V, and plot the resulting $R(V_g)$ traces in Fig. 2a. The photo-induced doping leads to a shift of CNP to the $V_g$ set points, corresponding to an n-doping concentration. Remarkably, the peaks of all $R(V_g)$ remain as sharp as the pristine sample. This is in striking contrast to doping induced by adsorbed atoms, where a significant increase of $R(V_g)$ peak width accompanies higher doping concentrations[18-20]. Figure 2b quantifies the charge transport properties by plotting the charge density fluctuations δ$n$ and mobilities μ close to the CNP for different photo-induced doping concentrations (see Supplementary Fig. 1). Evidently, the photo-induced doping preserves the excellent electrical transport properties of G/BN. The electron mobility μ remains almost constant over the entire doping concentrations. δ$n$ exhibits similar behavior, and increases only marginally while the doping level increased to $3\times10^{12}$cm$^{-2}$.

We demonstrate the flexibility of photo-induced doping to control the doping profile in G/BN optically in Fig. 2c. We create a high quality graphene *pn* junction by exposing one region of the sample to light with $V_g$ set at -20V (see inset of Fig. 2c). The resulting $R(V_g)$ trace taken with light off, is shown in Fig. 2c (blue line). Two distinct peaks are observed with similar heights that are separated by -16V. This transport behaviour is characteristic of a graphene *pn* junction[21-23]. This photo-induced doping can last for days if the device is kept in a dark environment, or it can be erased by exposing the sample to white light while $V_g$=0V. The erasure process usually requires 50 times higher illumination dosage (power density * exposure time) than the doping process. The red trace in Fig. 2c shows the $R(V_g)$ curve after the erasing process, and it exhibits a single sharp peak centred at $V_g$=0V as in the pristine device.

Combining the photo-induced modulation doping with photolithography techniques can enable scalable fabrication of high mobility graphene devices with arbitrary doping pattern. This fabrication scheme does not require any photoresist because G/BN itself is the photosensitive media. Additionally, the device is rewritable by controlling light illumination. It offers distinct advantages over alternative techniques in creating graphene *pn* junctions that require nontrivial multi-step fabrication process that reduces sample quality[21,22] and/or generates irreversible doping[24].

Next we study the microscopic processes responsible for the photo-induced modulation doping in G/BN. First, we need to identify the electronic states being excited by photons that lead to photo-doping effects. The initial optical excitation might take place at the G-BN interface (by exciting graphene or special interface states[12,13]) or inside BN. These two cases can be differentiated by examining the photo-doping dynamics with different BN thicknesses, because the photo-doping rate will have a positive correlation with BN thickness if electronic states in BN are excited, and it will be independent of BN thickness if interface states are excited. Figure 3a shows the experimental data for G/BN heterostructures with 20nm and 110nm BN flakes. These two BN flakes were exfoliated from the same scotch tape to ensure that they have the same physical properties. At *t*=0s we set $V_g$ = -30V for the 20nm BN sample (-40V for the 110nm BN sample to ensure the same doping density) and switch on the light. We observe that the dynamic resistance increase (or the photo-doping rate) is much faster in 110nm BN sample than that in 20nm BN sample under the same illumination condition. Results for more samples are plotted in the inset of Fig. 3a. We observe a systematic increase of photo-doping rate in thicker BN flakes, although a significant variation of the doping rate is observed. Our results suggest that optical excitation of electronic states inside BN initialises the photo-induced doping in G/BN heterostructures. The typical light power density we used is 0.1-1μW/μm$^2$, which is three orders magnitude lower than that required to observe photo-doping originated from interfacial charge traps in $SiO_2$[12,13].

Previously BN was always considered as an inert substrate because it has a bandgap of 6.4 eV[25]. Therefore, it is quite surprising that the photo-induced doping in G/BN originates from optical excitation of electronic states in BN. Obviously visible photons used in the experiment cannot excite the bandgap transitions, but they can excite defect states in BN. In Fig. 3b we illustrate a physical picture of the photo-induced doping in G/BN starting with defect states in BN. Graphene is initially hole-doped at a negative $V_g$ in the dark, and an electric field emanates from graphene to silicon. Upon optical illumination, electrons of donor-like defects in BN are excited by photons to the conduction band. These excited electrons can be mobile and move towards graphene following the existing electrical field and then enter graphene. The ionised defects are positively charged and localised in BN, and they effectively screen the backgate. The process continues until the electric field in BN vanishes and graphene becomes charge neutral, as we observed experimentally. Because the ionised defects are within the BN flake, which are on average tens of nanometers away from graphene, they introduce minimum extra scattering in graphene[26]. Additionally, correlations between these charged defects can reduce the scattering even more[27]. Negatively charged defects can also be generated in BN if optical excitation excites acceptor-like defects in BN at a positive $V_g$. However, the process is much slower in our experiment, presumably due to a much lower concentration of acceptor-like defects in BN.

With this microscopic understanding, we can employ the photo-induced doping in G/BN as a tool to study the nature of the BN defect states. The optical absorption cross-section of defect states in BN is proportional to the generation rate of ionised defects, which can be measured sensitively through their effect on graphene electrical transport (see Supplementary Fig.

5). Moreover, we can probe selectively the donor-like defects and acceptor-like defects by setting $V_g$ to be negative and positive, respectively. Figure 4 shows the photo-doping rate for acceptor-like defects (red trace) and donor-like defects (green trace) as a function of photon energy. We found that photo-doping rate, which is proportional to optical absorption of the defect states, varies with the defect type and the excitation energy. For donor-like states, we found that the absorption cross-section keeps increasing in the experimental spectral range (up to 2.6 eV). A similar trend is observed for acceptor-like states, but the photo-doping rate is approximately two orders of magnitude lower. These spectral dependences suggest deep donor and acceptor defect levels close to the middle of BN bandgap. Previous theoretical work found that the dominating donor and acceptor absorption resonance are from nitrogen vacancy at 2.8eV and carbon impurity (substituting a nitrogen atom) at 2.6eV[28]. These resonances are slightly beyond our experimental spectral range, but are consistent with the observed strong rise at below 2.6 eV. Greater spectrum range in the higher photon energy regime will be necessary to directly probe the absorption resonances from the defect levels.

In summary, we observed strong photo-induced modulation doping effect in G/BN heterostructures and provide a microscopic description of its origin. This effect can enable flexible fabrication of graphene devices through controlled light exposure as in photolithography. Moreover, it allows for repeatable writing and erasing of the doping features, and preserves the very high mobility of G/BN. This new and simple technique of creating inhomogeneous doping in a high mobility graphene device opens the door for novel scientific studies and applications.

**Method**

Most of the samples were fabricated with the transfer technique developed by P.J. Zomer et al[11] and using standard electron beam lithography. We employed an h-BN thickness of 10-110nm and a $SiO_2$ thickness of 285nm as the dielectrics for electrostatic gating. Monolayer graphene exfoliated from Kish graphite is deposited onto a Methyl methacrylate(MMA) polymer and transferred to exfoliated hexagonal boron nitride (h-BN) that is sitting on a $SiO_2$/Si wafer. We employed a standard AC current biased lock-in technique with 15 nA at 97.13 Hz in a cryostat at 77 K, and under a vacuum level of $10^{-5}$ Torr. A fibre-based supercontinuum laser is guided into the cryostat through an optical window and different wavelength is obtained by diffraction from a grating. We also investigated several devices using epitaxial grown graphene on BN flakes[29], and they exhibit similar behavior.

**Acknowledgements**

The authors thank P. Jarillo-Herrero and N. Gabor for stimulating discussions and Brian Standley for help with data acquisition software. Graphene synthesis, device fabrication, and theoretical analysis were supported by the Office of Naval Research (award N00014-13-1-0464). Optical and electrical measurements of this work were mainly supported by Office of Basic Energy Science, Department of Energy under contract No. DE-SC0003949 (Early Career Award) and DE-AC03-76SF0098 (Materials Science Division). F.W. acknowledges the support from a David and Lucile Packard fellowship. J.V.J. acknowledges the support from the UC President's Postdoctoral Fellowship.

**Author contributions**

F.W. and L.J. conceived the experiment. L.J. and J.V.J. carried out optical and electronic measurements, J.V.J., E.H., S.K., C.N., H.T, W.Y. contributed to sample fabrication, K.W. and T.T. synthesised the h-BN samples, F.W., J.V.J. and L.J. performed data analysis and theoretical analysis. F.W., L.J. and J.V.J. co-wrote the manuscript. All authors discussed the results and commented on the paper.

**Additional information**

The authors declare no competing financial interests. Supplementary information accompanies this paper at www.nature.com/naturenanotechnology. Reprints and permission information is available online at http://npg.nature.com/reprintsandpermissions/. Correspondence and requests for materials should be addressed to F.W.

**Figure 1: Experimental observation of photo-induced modulation doping effect in G/BN heterostructures.** (a) Representative $R(V_g)$ data in a G/BN device showing a sharp resistance peak at $V_g=0$ without light exposure. Inset is an optical micrograph of device, where graphene (outlined by the solid line) is on a BN flake (blue, 18nm thick) and contacted by chrome/gold electrodes. The scale bar represents 8μm. (b) $R(V_g)$ trace as $V_g$ sweeps from 70V to -70V and then back to 70V with the device exposed to light. Light illumination changed the gating behavior in graphene, where negative gating becomes ineffective and the graphene resistance remains at the CNP value. The BN flake is ~100nm thick and the same for c. (c) $R(V_g)$ trace when $V_g$ sweeps from 20V to -50V and the light illumination is switched on and off alternately. Graphene resistance is pinned at charge neutral point value whenever light is switched on at negative $V_g$, and shows normal gating behavior when light is switched off. (d) and (e) illustrate the charge distribution in G/BN device when light is switched off and on, respectively. Positive charges are accumulated in BN under light illumination at negative $V_g$, which effectively screens the backgate and keeps graphene at CNP.

**Figure 2: Transport characteristics of G/BN after photo-induced modulation doping.** (a) $R(V_g)$ traces displaying high mobility charge transport in G/BN devices with photo-induced modulation doping. The red trace shows the behavior of a pristine sample. The other traces were acquired after photo-doping where the graphene CNP has been set at $V_g$=-10, -20, -30, -40 and -50V, respectively (from right to left). The BN flake is ~80nm thick. (b) Quantitative determination of electron mobility $\mu$ (squares, left axis) and charge density fluctuation $\delta n$ at CNP (triangles, right axis) from $R(V_g)$ traces in (a) at different photo-induced doping density $n_{PD}$. The error bars represent +/- one standard deviation. (c) Generation and erasure of a *pn* junction in G/BN heterostructure with light. An inhomogeneous photo-doping can be established by illuminating part of G/BN device with light at $V_g$=-20V (inset), which results in an $R(V_g)$ response typical of a graphene *pn* junction (blue trace). Subsequent exposure of the device to light at $V_g$=0V erases the inhomogeneous doping and recovers the $R(V_g)$ response of pristine graphene (red trace).

**Figure 3: Dynamics and origin of photo-induced modulation doping effect.** (a) Photo-induced doping density as a function of time in two devices under the same illumination condition and gating electrical field. The black and red traces correspond to the BN flake thickness of 20nm and 110nm, respectively. Evidently, the rate of photo-doping is significantly higher in the G/BN device with a thicker BN flake. Inset: Photo-doping rate in more samples show that photo-doping rate increases systematically with the BN thickness, although it has significant fluctuation from sample to sample. See section 4 in supplementary information for the definition of error bars. (b) An illustration of the photo-doping mechanism, where optical excitation first excites electrons from defects in BN. The excited electrons move into graphene under the applied gate electrical field, and the positively charged defects lead to the modulation doping in graphene when the light and $V_g$ is off.

**Figure 4: Optical spectrum of defect states in h-BN.** Photo-doping rate as a function of photon energy for donor (acceptor) states is extracted by applying a negative (positive) $V_g$. The green trace shows that the photo-doping rate (and therefore the absorption cross-section) from donor states keeps increasing in the experimental spectral range (up to 2.6 eV). A similar trend is observed for acceptor-like states (red trace), but this photo-doping rate is about two orders of magnitude lower. These spectral dependences suggest deep donor and acceptor defect levels close to the middle of BN bandgap. The BN flake is ~60nm thick.

**Figure 1: Experimental observations and illustration of photo-induced modulation doping effect**

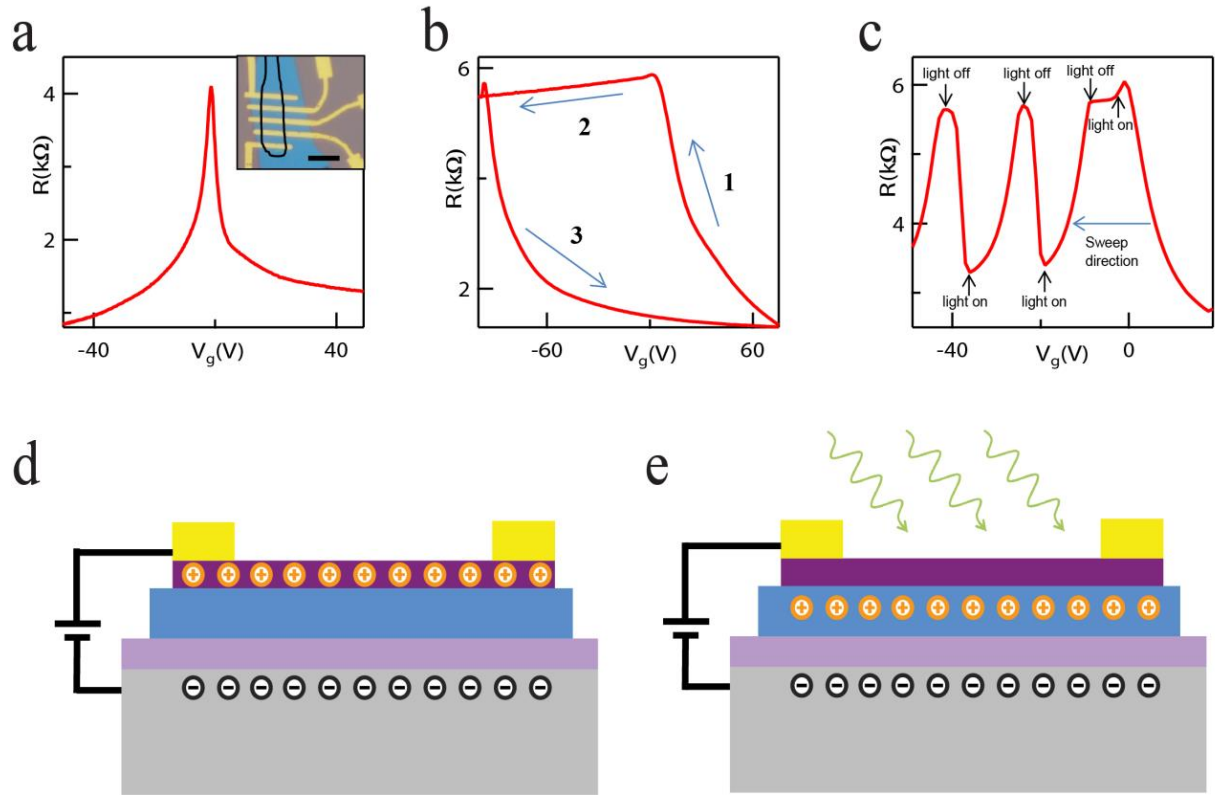

**Figure 2: Transport characteristics of G/BN after photo-induced modulation doping**

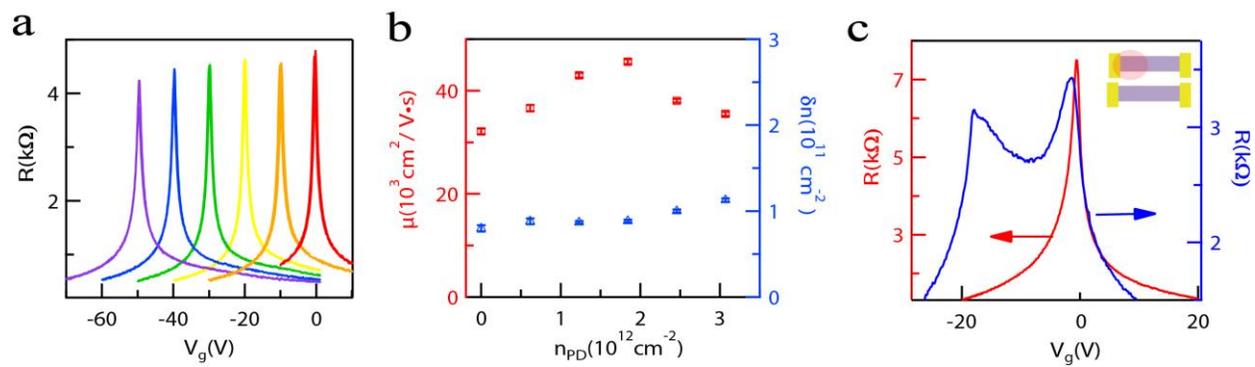

**Figure 3: Dynamics and origin of photo-induced modulation doping effect**

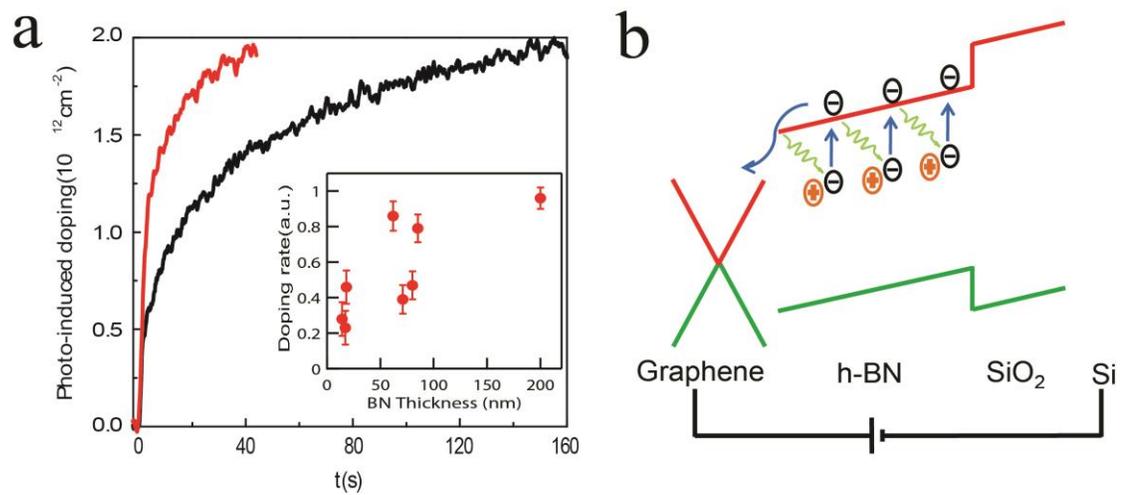

**Figure 4: Optical spectrum of defect states in h-BN**

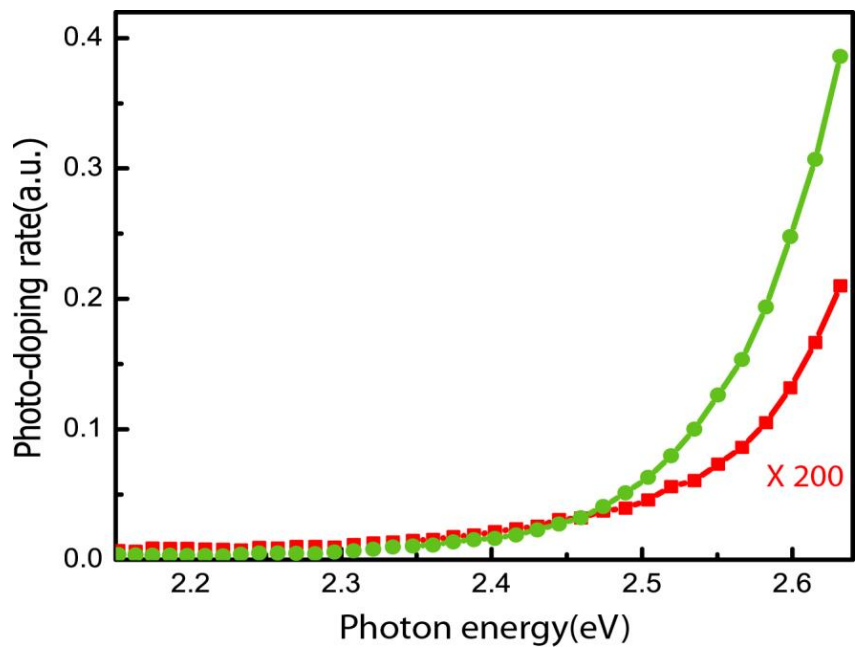